\documentclass[12pt]{article}
\pdfoutput=1
\usepackage{jheppub}
\usepackage{graphicx,epsfig,amsmath,amssymb}

\def\nn{\nonumber}

\def\be{\begin{equation}}
\def\ee{\end{equation}}
\def\bea{\begin{eqnarray}}
\def\eea{\end{eqnarray}}

\newcommand{\mhh}{m_{hh}}

\newcommand{\pth}{p_{T,h}}
\newcommand{\ct}{c_t}
\newcommand{\ctt}{c_{tt}}
\newcommand{\chhh}{c_{hhh}}
\newcommand{\cg}{c_{ggh}}
\newcommand{\cgg}{c_{gghh}}
\title{Exploring anomalous couplings in Higgs boson pair production through shape analysis}

\author[a]{M.~Capozi,}
\author[a]{G.~Heinrich}

\affiliation[a]{Max Planck Institute for Physics, F\"ohringer Ring 6, 80805 M\"unchen, Germany}

\emailAdd{mcapozi@mpp.mpg.de}
\emailAdd{gudrun@mpp.mpg.de}

\preprint{{\small  MPP-2019-183}}

\abstract{
We classify shapes of Higgs boson pair invariant mass distributions $m_{hh}$,
calculated at NLO with full top quark mass dependence, and visualise how distinct classes of shapes relate to the underlying
coupling parameter space.
Our study is based on a five-dimensional parameter space relevant for Higgs boson pair production in a non-linear Effective Field Theory framework.
We use two approaches: an analysis based on predefined shape types and a classification into shape clusters based on unsupervised learning. 
We find that our method based on unsupervised learning is able to capture shape features very well and therefore allows a more detailed 
study of the impact of anomalous couplings on the $m_{hh}$ shape
compared to more conventional approaches to a shape analysis.
}

\begin{document}

\maketitle
\newpage


\section{Introduction}

The Higgs sector as we see it today is probably just a glimpse of an underlying more general structure still awaiting to be explored. 
Manifestations of new physics at higher scales would lead to operators which on one hand introduce new, effective couplings and on the other hand also modify interactions known in the Standard Model (SM).
Therefore it is a primary goal for collider physics in the next decades to constrain the couplings, in particular in the Higgs sector, to an unprecedented precision.
This is particularly true for the Higgs boson self couplings, in order to find out whether the Higgs potential is indeed of the form assumed by the SM. Deviations from this form could provide strong hints about how to extend the SM.

The trilinear Higgs boson coupling $\lambda$ can be constrained by measurements of Higgs boson pair production~\cite{Aad:2019uzh,Sirunyan:2018two}, where the gluon fusion channel yields the largest cross section, and the most stringent 95\% CL limits
on the total $gg\to HH$ cross section at $\sqrt{s}=13$\,TeV are currently $\sigma^{HH}_{\rm{max}}= 6.9\times \sigma_{SM}$, constraining trilinear coupling modifications to the range $-5.0\leq \lambda/\lambda_{\rm{SM}}\leq 12.0$~\cite{Aad:2019uzh}.

The trilinear Higgs couplings can also be constrained in an indirect way, 
through measurements of processes which are sensitive to these couplings via electroweak corrections~\cite{McCullough:2013rea,Gorbahn:2016uoy,Degrassi:2016wml,Bizon:2016wgr,Maltoni:2017ims,Kribs:2017znd,Degrassi:2017ucl,Nakamura:2018bli,Kilian:2018bhs,Maltoni:2018ttu,Vryonidou:2018eyv,Gorbahn:2019lwq}.
Such processes offer important complementary information, however they are susceptible to other BSM couplings entering the loop corrections 
at the same level, and therefore the limits on $\chhh=\lambda/\lambda_{\rm{SM}}$ extracted this way may be more model dependent than the ones extracted from the direct production of Higgs boson pairs.
A corresponding experimental analysis based on single Higgs boson production processes has been performed~\cite{ATL-PHYS-PUB-2019-009}, and recently combined constraints from single and double Higgs boson production became available~\cite{ATLAS:2019pbo}.
The idea of indirect constraints through loop corrections also has been employed trying to constrain the quartic Higgs boson self-coupling from (partial) EW corrections to Higgs boson pair production~\cite{Bizon:2018syu,Borowka:2018pxx}.


Theoretical constraints on $\chhh$ are rather loose if derived in a largely model independent way. 
Recent work based on general concepts like vacuum stability and perturbative unitarity suggests that $|\chhh|\lesssim 4$ for a new physics scale in the few TeV range~\cite{Falkowski:2019tft,Chang:2019vez,DiLuzio:2017tfn,DiVita:2017eyz}.
More specific models can lead to more stringent bounds, see e.g. Refs.~\cite{Braathen:2019pxr,Basler:2018dac,Babu:2018uik,Adhikary:2017jtu,Lewis:2017dme}.
Recent phenomenological studies about the precision that could be reached for the trilinear coupling at the (HL-)LHC and future hadron colliders are summarised in Refs.~\cite{Dawson:2018dcd,Cepeda:2019klc,DiMicco:2019ngk}.

\medskip

Higgs boson pair production in gluon fusion in the SM has been calculated at leading order in Refs.~\cite{Eboli:1987dy,Glover:1987nx,Plehn:1996wb}, and at NLO in the $m_t\to\infty$ limit, rescaled with the full Born matrix element,  in Ref.~\cite{Dawson:1998py}. Ref.~\cite{Maltoni:2014eza} contains the full top quark mass dependence in the real radiation, 
while the virtual part is calculated in the heavy top limit.
The NLO QCD corrections with full top quark mass dependence became available more recently~\cite{Borowka:2016ehy,Borowka:2016ypz,Baglio:2018lrj,Baglio:2020ini}.
Implementations of the full NLO QCD corrections in parton shower Monte Carlo programs are also available~\cite{Heinrich:2017kxx,Jones:2017giv,Heinrich:2019bkc}.

NNLO QCD corrections have been computed in the $m_t\to\infty$ limit in Refs.~\cite{deFlorian:2013uza,deFlorian:2013jea,Grigo:2014jma,Grigo:2015dia,deFlorian:2016uhr}. 
 The calculation of Ref.~\cite{deFlorian:2016uhr} has been combined with results including the top quark mass dependence as far as available in Ref.~\cite{Grazzini:2018bsd}, and the latter has been supplemented by soft gluon resummation in Ref.~\cite{deFlorian:2018tah}. 
The scale uncertainties at NLO are still at the 10\% level, while they are decreased to about 5\% when including the NNLO corrections.
The uncertainties due to the chosen top mass scheme have been assessed in Ref.~\cite{Baglio:2018lrj}. 

Analytic approximations for the top quark mass dependence of the two-loop amplitudes in the NLO calculation have been studied in Refs.~\cite{Grober:2017uho,Bonciani:2018omm,Xu:2018eos,Davies:2018ood,Davies:2019xzc}. Complete analytic results in the high energy limit have been presented in Ref.~\cite{Davies:2018qvx}; the latter have been combined with the full NLO results in the regions where they are more appropriate in Ref.~\cite{Davies:2019dfy}.

The effects of operators within an Effective Field Theory (EFT) description of Higgs boson pair production have been studied at LO QCD in Refs.~\cite{Contino:2012xk,Goertz:2014qta,Chen:2014xra,Azatov:2015oxa,Dawson:2015oha,Carvalho:2015ttv,Cao:2015oaa,Cao:2016zob,DiVita:2017eyz,deBlas:2018tjm} and at NLO in the $m_t\to\infty$ limit in Refs.~\cite{Grober:2015cwa,Grober:2016wmf,Maltoni:2016yxb}, including also CP-violating operators~\cite{Grober:2017gut}. EFT studies at NNLO in the $m_t\to\infty$ limit are also available~\cite{deFlorian:2017qfk}.
In Ref.~\cite{Buchalla:2018yce} for the first time the full NLO QCD corrections have been combined with an EFT approach to study BSM effects.

It is well known that Higgs boson pair production in gluon fusion is a process where delicate cancellations occur between contributions containing the trilinear Higgs coupling and box-type contributions not containing the trilinear coupling.
While the destructive interference between these contributions is usually seen as a curse leading to small cross sections, it can be turned into a virtue when analyzing the shapes of distributions, as for example the di-Higgs invariant mass distribution $\mhh$, because even small anomalous couplings can lead to characteristic shape changes. Therefore it is important to investigate in which way the shapes are influenced by a certain 
configuration in the coupling parameter space. 

The idea of a shape analysis has been pursued already in various ways based on LO studies, see e.g.  Refs.~\cite{Chen:2014xra, Azatov:2015oxa, Dawson:2015oha,Carvalho:2015ttv,Carvalho:2016rys,Carvalho:2017vnu}.
In Ref.~\cite{Carvalho:2015ttv}, a cluster analysis is proposed to define 12 benchmark points in a 5-dimensional non-linear EFT parameter space which result from clusters of ``similar" shapes.
 The similarity measure in this case is based on a binned likelihood ratio using LO predictions for the observables $\mhh$, $\cos\theta^*$ and $\pth$.
 In Ref.~\cite{Buchalla:2018yce} it was analyzed how the $\mhh$ and $\pth$ distributions change when going from LO to NLO for the benchmark points defined in Ref.~\cite{Carvalho:2015ttv}.

As a function of the 5-dimensional coupling parameter space, the $\mhh$ distribution can have a few characterising features, such as an enhanced low-$\mhh$ region, a double peak, a single peak or an enhanced tail. 
Some of these features can be attributed rather easily to a certain anomalous coupling, 
for example, an enhanced low-$\mhh$ region is naturally produced by large values of $|\chhh|$.
Other features of the $\mhh$-shape, like a double peak or a SM-like distribution, are harder to attribute to a certain coupling configuration, as there are a multitude of configurations leading to such shapes.
This is also reflected in the cluster analysis proposed in Ref.~\cite{Carvalho:2016rys}, where (a) very different coupling configurations can end up in the same cluster, and (b) the same cluster can contain shapes which ``by eye" look quite different (for example ``double peak" and ``single peak").

Therefore it is desirable to seek for alternative methods to extract information about the underlying parameter space from the shape of distributions in Higgs boson pair production.
In this work we first classify the shapes of Higgs boson pair invariant mass distributions, calculated at full NLO, into four characteristic types. 
We visualise the underlying 5-dimensional EFT parameter space producing these shape types, projecting onto 2-dimensional subspaces. We also comment on the shape of the $\pth$ distribution.
Then we refine the shape analysis, applying an unsupervised learning algorithm based on an autoencoder to identify patterns in the shapes of the $\mhh$ distribution. 
We use the {\tt KMeans} clustering algorithm from {\tt scikit-learn}~\cite{scikit} and ask for a classification of the shapes into a certain number of clusters. 
One aim of this study is to offer an alternative to the cluster analysis proposed in Refs.~\cite{Carvalho:2015ttv, Carvalho:2016rys, Carvalho:2017vnu} and earlier work, another aim is to provide an analysis based on full NLO results.
The present study allows us to associate certain shapes more globally with distinct regions in the parameter space, in this sense going beyond a benchmark point analysis.
Nonetheless, to facilitate a future more quantitative analysis, for example a profile likelihood study, we identify new benchmark points, based on the cluster centers given by our procedure. 

The application of machine learning techniques in high energy physics, in particular to constrain the EFT/new physics parameter space, 
has been brought forward already some time ago~\cite{deOliveira:2015xxd,Brehmer:2016nyr,Brehmer:2018kdj,Brehmer:2018eca}.
There are also successful applications in jet and top quark identification~\cite{Guest:2016iqz,Kasieczka:2017nvn,Datta:2017lxt,Louppe:2017ipp,Larkoski:2017jix,Macaluso:2018tck,Bollweg:2019skg,Kasieczka:2019dbj,Butter:2019cae,Moreno:2019bmu,Chen:2019uar,Chang:2019ncg} and PDF fits~\cite{Carrazza:2019mzf}. 
Machine learning in new physics searches is mostly used for anomaly detection~\cite{Hajer:2018kqm,DeSimone:2018efk,Andreassen:2018apy,DAgnolo:2018cun} and to improve the sensitivity to new physics, optimising the signal to background ratio~\cite{Chang:2017kvc,Brehmer:2018kdj,Brehmer:2018eca,Englert:2018cfo,Blance:2019ibf,Brehmer:2019gmn,Freitas:2019hbk}. 

The remainder of this work is structured as follows: In Section \ref{sec:procedure} we explain the framework our data samples are based on. We define four different shape types for the $\mhh$ distribution and 
 visualise the parameter space underlying the predefined shape types. In Section \ref{sec:unsupervised}
we describe our cluster analysis based on unsupervised learning and show how the clusters found by this procedure relate to the underlying parameter space. We also definine seven new benchmark points, before we conclude.

\section{Classification through predefined shape types}
\label{sec:procedure}

\subsection{Parametrisation of anomalous couplings in the Higgs sector}
\label{sec:supervised}

As a starting point we use the effective Lagrangian in a non-linear Effective Field Theory (``Higgs Effective Field Theory, HEFT") relevant for 
Higgs boson pair production, assuming CP conservation, up to order 4 in the chiral expansion~\cite{Buchalla:2015wfa,Buchalla:2018yce}:
\begin{align}
{\cal L}\supset 
-m_t\left(c_t\frac{h}{v}+c_{tt}\frac{h^2}{v^2}\right)\,\bar{t}\,t -
c_{hhh} \frac{m_h^2}{2v} h^3+\frac{\alpha_s}{8\pi} \left( c_{ggh} \frac{h}{v}+
c_{gghh}\frac{h^2}{v^2}  \right)\, G^a_{\mu \nu} G^{a,\mu \nu}\;.
\label{eq:ewchl}
\end{align}
In the SM $c_t=c_{hhh}=1$ and $c_{tt}=c_{ggh}=c_{gghh}=0$.
The chromomagnetic operator is absent in (\ref{eq:ewchl}) because it 
contributes to $gg\to hh$ only at higher order in the chiral counting.
The coefficients $c_{ggh}$ and $c_{gghh}$ are related in SMEFT (``SM Effective Field Theory")~\cite{Berthier:2015oma,Brivio:2017btx,DiMicco:2019ngk}, however in HEFT there is not necessarily a relation between the two parameters.
To clarify the relation to the widely used SMEFT operators, we  briefly comment on the SMEFT Lagrangian here.
The dimension-6 terms relevant for $gg\to hh$ can be written as~\cite{Grober:2015cwa,Azatov:2015oxa}
\begin{align}
\Delta{\cal L}_6 &=
\frac{\bar c_H}{2 v^2}\partial_\mu(\phi^\dagger\phi)\partial^\mu(\phi^\dagger\phi)
+\frac{\bar c_u}{v^2} y_t(\phi^\dagger\phi\, \bar q_L\tilde\phi t_R +{\rm h.c.})
-\frac{\bar c_6}{2 v^2}\frac{m^2_h}{v^2} (\phi^\dagger\phi)^3
\nonumber\\
&+\frac{\bar c_{ug}}{v^2} g_s
(\bar q_L\sigma^{\mu\nu}G_{\mu\nu}\tilde\phi t_R +{\rm h.c.})
+\frac{4\bar c_{g}}{v^2} g^2_s \phi^\dagger\phi\, G^a_{\mu\nu}G^{a\mu\nu}\;.
\label{lsmeft}
\end{align}
Relating the coefficients $\bar c_i$ in Eq.~(\ref{lsmeft}) to the couplings of 
the physical Higgs field $h$ and comparing with the corresponding parameters 
of the chiral Lagrangian Eq.~(\ref{eq:ewchl}), one finds, 
after a field redefinition of $h$ to eliminate $\bar c_H$ from the
kinetic term,
\begin{equation}\label{cthhh}
c_t=1-\frac{\bar c_H}{2}-\bar c_u\, ,\quad
c_{tt}=-\frac{\bar c_H + 3\bar c_u}{2}\, ,\quad
c_{hhh}=1-\frac{3}{2}\bar c_H +\bar c_6\;,
\end{equation}
\begin{equation}\label{cghh}
c_{ggh}=2 c_{gghh}=128\pi^2\bar c_g\;.
\end{equation}
Note that, assuming an underlying weakly coupled gauge theory, 
dimension-6 operators involving field-strength tensors can only be 
generated through loop diagrams~\cite{Arzt:1994gp}.
Their coefficients then come with an extra factor of $1/16\pi^2$. 
In this case, the coefficients $\bar c_{ug}$ and $\bar c_g$ in Eq.~(\ref{lsmeft}) are counted
as order $(1/16\pi^2)(v^2/\Lambda^2)$, while $\bar c_H$, $\bar c_u$ and
$\bar c_6$ are still of order $v^2/\Lambda^2$.
For more details about the difference between HEFT and SMEFT we refer to Refs.~\cite{Buchalla:2018yce,DiMicco:2019ngk}.

We produce our data  using the differential distributions calculated in Ref.~\cite{Buchalla:2018yce}, parametrised in terms of coefficients $A_i$ for each coupling combination occurring in the (differential) NLO cross section, which allows for a fast evaluation:
\begin{align}
\frac{d \sigma}{dm_{hh}} =& A_1c_{t}^4+A_2c_{tt}^2+A_3c_{t}^2c_{hhh}^2+A_4c_{ghh}^2c_{hhh}^2 + A_5c_{gghh}^2 + A_6c_{tt}c_{t}^2+A_7c_{t}^3c_{hhh} \nn\\ &+A_8c_{tt}c_{t}c_{hhh}+A_9c_{tt}c_{ggh}c_{hhh}+A_{10}c_{tt}c_{cgghh}+A_{11}c_{t}^2c_{ggh}c_{hhh}+A_{12}c_{t}^2c_{gghh} \nn\\ & + A_{13}c_{t}c_{hhh}^2c_{ghh}+A_{14}c_{t}c_{hhh}c_{gghh}+A_{15}c_{ggh}c_{hhh}c_{gghh} + A_{16}c_{t}^3c_{ggh}+A_{17}c_{t}c_{tt}c_{ggh} \nn\\ & +A_{18}c_{t}c_{ggh}^2c_{hhh}+A_{19}c_{t}c_{ggh}c_{gghh}+A_{20}c_{t}^2c_{ggh}^2+A_{21}c_{tt}c_{ggh}^2+A_{22}c_{ggh}^3c_{hhh}\nn\\
&+A_{23}c_{ggh}^2c_{gghh} \;.
\label{eq:Ai_mhh}
 \end{align}
The coefficients $A_i$ are evaluated in bins of width 20\,GeV  from 250\,GeV to 1050\,GeV, i.e. for 40 bins. They are available with Ref.~\cite{Buchalla:2018yce} as {\tt  .csv} tables, in units of fb/GeV, for $\sqrt{s}=13, 14$ and 27\,TeV.
The median of the statistical uncertainties of the differential coefficients $A_i$ does not exceed 3\%, however in the bins beyond $\mhh\gtrsim 650$\,GeV some $A_i$ coefficients have uncertainties in the 20-30\% range.

%

\subsection{Definition of shape types}
\label{sec:shapes}
\begin{figure}
\includegraphics[width=\linewidth]{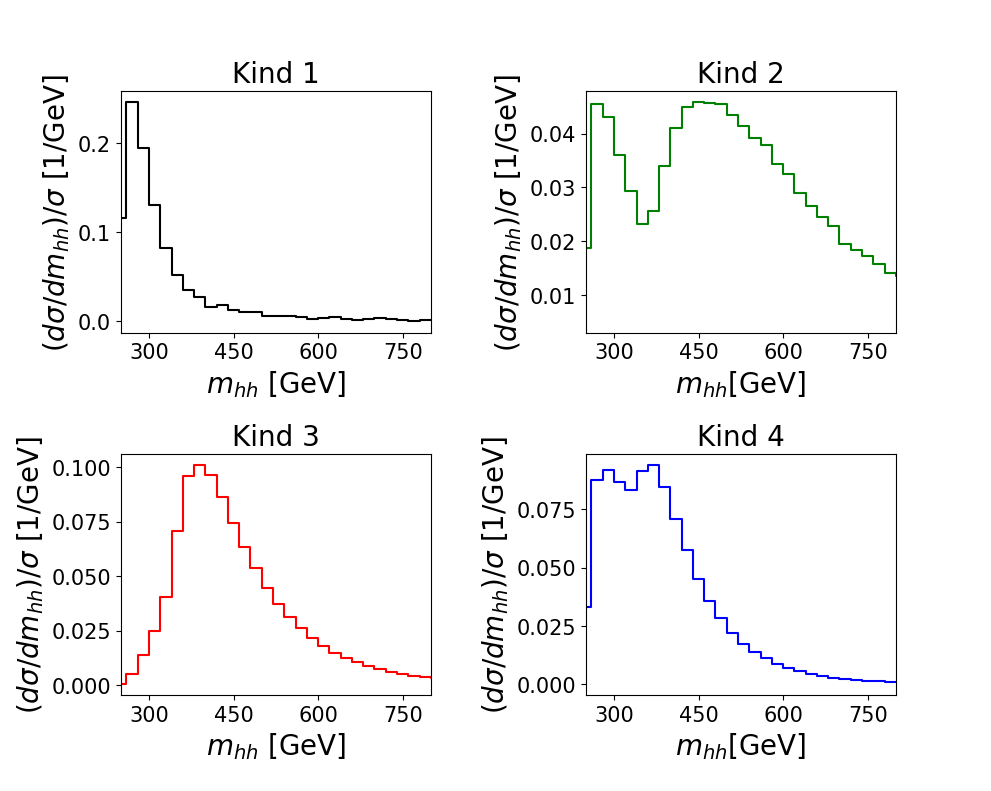}
  \caption{The four kinds of shapes defined in our analyzer to classify the $\mhh$ distributions. The colours correspond to the colours shown in Figs.~\ref{fig:HM1} to \ref{fig:HM5}.}
  \label{fig:4k}
\end{figure}

We distinguish four types of characteristic shapes for the Higgs boson invariant mass distribution $\mhh$:
\begin{enumerate}
\item Enhanced low $\mhh$ region, constantly falling distribution as $\mhh$ increases.
\item Double peak with peaks separated by more than 100 GeV.
\item Single peak  near the $t\bar{t}$ production threshold at $\mhh\sim 346$\,GeV.
\item Double peak with peaks separated by less than 100 GeV.
\end{enumerate}
Examples of the four shape types are shown in Fig.~\ref{fig:4k}. 
According to our classification the Standard Model shape is contained in distributions of kind 3.  
Certainly there is some arbitrariness in the definition of these shapes. 
For example, shapes of kind 4 would move to kind 1 or 3 for bin widths $\geq 100$\,GeV. However, the other three shape types are quite robust and would be clearly distinguishable experimentally.

Based on the parametrisation in Eq.~(\ref{eq:Ai_mhh}), the normalised differential cross section is computed for a 5-dimensional grid in the coupling parameter space and according to its behaviour is classified into one of the four shape types. For this purpose we wrote a function, called ``analyzer" in the following, that checks the slopes of the distribution and puts it into the corresponding class. 
At this stage the shape classes are mutually exclusive.
For each point in the coupling parameter space, we also consider the variations of the result in each bin due to inclusion of the  statistical uncertainties on the coefficients $A_i$.
If the shape obtained after these variations belongs to  a different kind, we exclude that point from the data set. 
We find that for shape type 4 about 20\%  of points fall into this category and are therefore excluded, while for shape type 2 it is about 8\%, and for types 1 and 3 it is less than 5\%.
Scale variations have not been included, as they tend to be rather uniform over the whole $\mhh$ range~\cite{Buchalla:2018yce,Heinrich:2019bkc} and therefore would not significantly modify our shape analysis.


\subsection{Classification of $\mhh$ distributions} 
\label{sec:mhh_analysis}

Our results for the $gg\to hh$ cross sections at NLO are produced for a centre-of-mass energy of $\sqrt{s}=13$\,TeV,  using PDF4LHC15$\_$nlo$\_$100$\_$pdfas~\cite{Butterworth:2015oua} parton distribution functions interfaced via LHAPDF, along with the corresponding value for $\alpha_s$.
The masses have been set to $m_h=125$\,GeV, $m_t=173$\,GeV and the top quark width has been set to zero.

\begin{figure}[htb]
\centering
\includegraphics[width=0.8\linewidth]{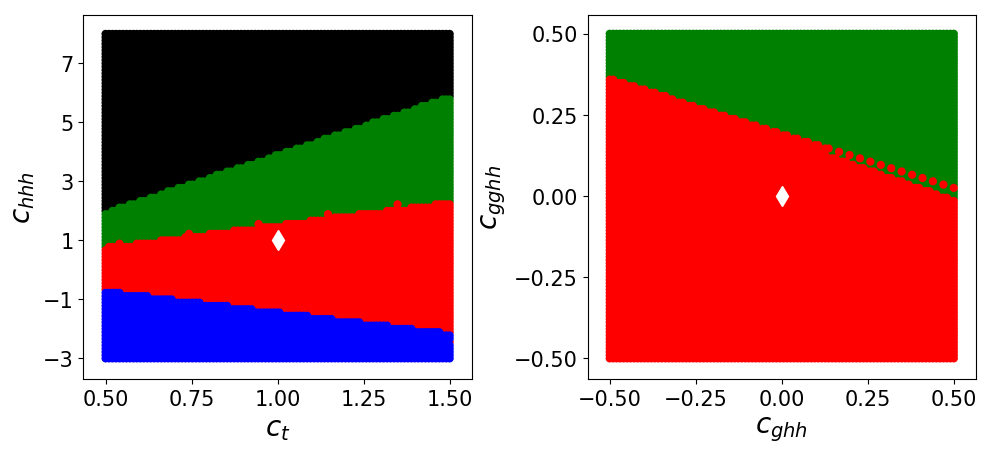}
\caption{The parameter regions leading to each predefined shape type in the $c_t-c_{hhh}$ (left) and $c_{ggh}-c_{gghh}$ (right) parameter spaces.
The black area denotes shapes of kind 1 (enhanced low $\mhh$ region; green: kind 2 (well separated double peaks), red: kind 3 (SM-like), blue: kind 4 (close-by double peaks).
The white diamonds mark the Standard Model point.}
  \label{fig:HM1}
\end{figure}

\begin{figure}[htb]
\centering
\includegraphics[width=0.8\linewidth]{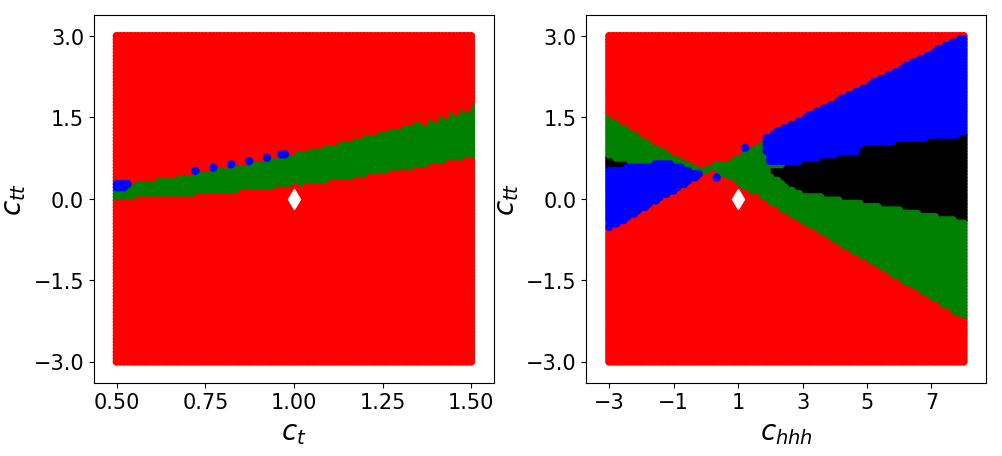}
\caption{The parameter regions associated to each shape type  in the $c_{t}-c_{tt}$ (left)  and $c_{hhh}-c_{tt}$ (right) parameter spaces.
For the colour code we refer to Fig.~\ref{fig:4k}.}
  \label{fig:HM2}
\end{figure}

\begin{figure}[htb]
\centering
\includegraphics[width=0.8\linewidth]{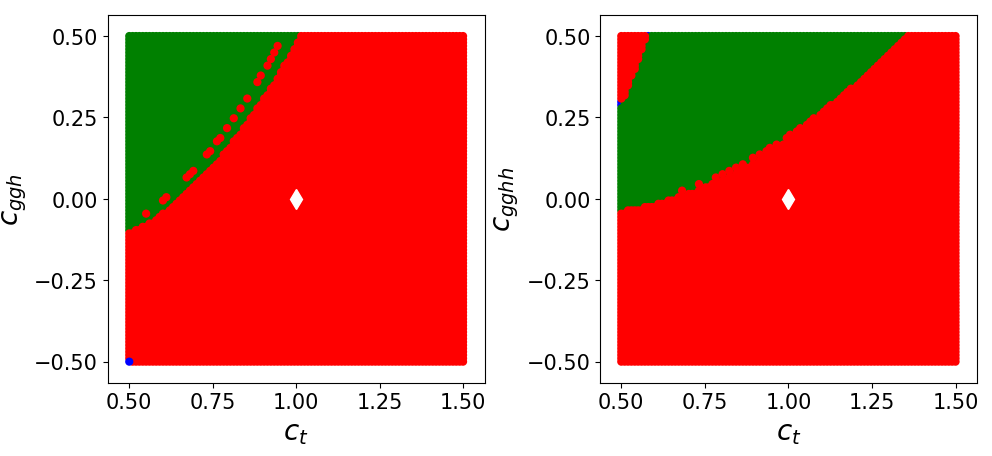}
  \caption{The parameter regions associated to each shape type in the $c_{t}-c_{ggh}$ (left) and   $c_{t}-c_{gghh}$ (right) planes.}
  \label{fig:HM3}
\end{figure}

\begin{figure}[htb]
\centering
\includegraphics[width=0.8\linewidth]{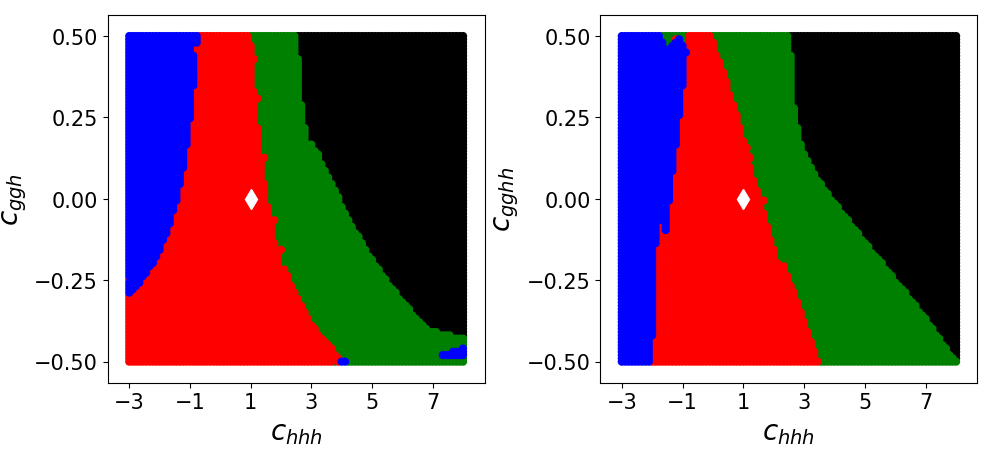}
  \caption{The parameter regions associated to each shape type in the $\chhh-c_{ggh}$ (left) and   $\chhh-c_{gghh}$ (right) planes.}
  \label{fig:HM4}
\end{figure}

\begin{figure}[htb]
\centering
\includegraphics[width=0.8\linewidth]{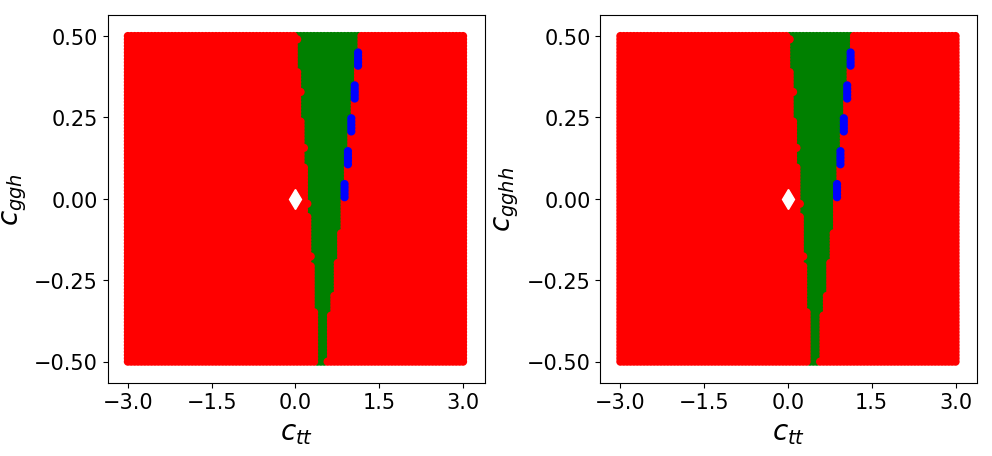}
  \caption{The parameter regions associated to each shape type in the $\ctt-\cg$ (left) and $\ctt-\cgg$ (right) planes.}
  \label{fig:HM5}
\end{figure}

We study the differential cross section as a function of five anomalous couplings, varying them in the ranges specified below,
\be
 \ct \in [0.5,1.5], c_{hhh} \in [-3,8], c_{tt} \in [-3,3], c_{ggh},c_{gghh} \in [-0.5,0.5]\;.
 \label{eq:ranges}
\ee
The ranges are motivated by current experimental constraints. For $\chhh$ we use a smaller range than suggested by experiment in order to focus more on the range where interesting shape features are present.
In order to visualise the results, we project out 2-dimensional slices of the 5-dimensional parameter space, fixing 
the other three couplings to their SM values. This leads to a total of ten configurations.
For each of these ten projections we generated a set of $10^6$ parameter pairs.
Feeding them through our analyzer we obtain the shape type produced by the given point in the coupling parameter space.
The results are shown in Figs.~\ref{fig:HM1}--\ref{fig:HM5}.
The white diamonds  denote the Standard Model  point in the parameter space.
Scale variations are not included, as they are rather uniform
over the whole $\mhh$ range~\cite{Buchalla:2018yce,Heinrich:2019bkc}
and therefore would not modify our shape analysis significantly.

In Fig.~\ref{fig:HM1} we display variations of the top quark Yukawa coupling  $c_t$ versus the trilinear Higgs coupling $c_{hhh}$ (left) and 
the effective gluon-Higgs couplings, $\cg$ versus $\cgg$ (right). In all the figures where two couplings are varied, the other three couplings are set to their SM values.   
It can be clearly seen that the shapes of kind 1, i.e. shapes with an enhanced low $\mhh$ region (marked in black), are resulting from larger $c_{hhh}$ values. The total  cross section as a function of $c_{hhh}$ is a parabola with a minimum around $c_{hhh} \approx 2.4$, while for $|c_{hhh}| \gtrsim 3$ and $\ct=1$ the distribution is enhanced in the low $\mhh$ region, where the triangle-type contributions dominate.
Larger/smaller values of $\ct$ shift this behaviour towards larger/smaller values of $\chhh$ because they enhance/decrease the box-type contributions.
For shapes of kind 2, i.e double peaks with a separation of more than 100\,GeV  (green), we find that such a shape can be produced for coupling values which are rather close to the SM values.
Shapes of kind 3 (red) are SM-like. They only cover about one quarter of the $\ct-\chhh$ plane.
Shapes of kind 4 (blue) have a double peak separated by less than 100\,GeV. For $\ctt=\cg=\cgg=0$, such structures only occur for negative values of $\chhh$, over the whole allowed $\ct$ range.

Considering variations of $c_{ggh}$ versus $c_{gghh}$, shown in the right-hand panel of Fig.~\ref{fig:HM1}, we find only shapes of kind 2 (green) and SM-like shapes (red). 
The existence of kind 2 shapes means that a double peak structure could be produced solely by effective Higgs-gluon couplings, while keeping $\chhh,\ct$ and $\ctt$ at their SM values. However, for the more likely case that $\cg$ deviates only slightly from zero~\cite{Sirunyan:2018sgc}, and so does $\cgg$, these couplings do not distort the SM shape significantly.

Variations of $c_t$ versus $c_{tt}$ and $\chhh$ versus $\ctt$ are shown in  Fig.~\ref{fig:HM2}.  Varying only $c_t$ and $c_{tt}$,  the shapes remain mainly SM-like. 
A small area in the $\ct-\ctt$ plane however contains doubly peaked $\mhh$ distributions,
which thus can originate from anomalous top-Higgs couplings only, while the trilinear Higgs coupling remains fixed at its SM value.

Turning to $c_{hhh}$ versus $c_{tt}$, displayed in the right-hand panel of Fig.~\ref{fig:HM2}, we find that 
for kind 1 and kind 4 shapes the parameter regions are split into two disconnected parts.
While shapes of kind 1 are favoured by large values of $\chhh$, it becomes clear that large values of $\ctt$, also related to triangle-type diagrams, can counterbalance this effect, 
because the top right corner is not a parameter region producing shapes of kind 1. If both $\chhh$ and $\ctt$ are large, it is more likely to produce a double peak structure with close-by peaks (kind 4, blue).
Further we see that shapes of kind 2 (well separated double peak structure, green) can be produced by values of $\ctt$ and $\chhh$ which are rather close to the SM values.

Fig.~\ref{fig:HM3} shows  variations of $\ct$ versus $\cg$ (left) and $\ct$ versus $\cgg$ (right). 
The  parameter space is dominated by SM-like shapes (red), however double peaks can occur as well (green).
We also see that $c_{gghh}$ acts similarly to $c_{ggh}$ in what concerns the shape.
%

For variations of  $c_{hhh}$ versus $c_{ggh}$, shown in Fig.~\ref{fig:HM4} (left), all four shape types can occur. 
The parameter region related to kind 1 (enhanced low $\mhh$, black) is at high values of $c_{hhh}$ as expected, and the kind 2 shapes (well separated double peak, green) can be seen as a transition from kind 1 to kind 3. Close-by double peaks (kind 4, blue) however are mostly associated to negative $\chhh$ values. Note that a similar pattern can be found in Fig.~\ref{fig:HM1} (left). 
%
Variations of  $c_{hhh}$ versus $c_{gghh}$, shown in  Fig.~\ref{fig:HM4} (right), are  similar in the overall behaviour, and again show that $c_{gghh}$ and $c_{ggh}$ have a similar impact on the shape.

Fig.~\ref{fig:HM5} shows variations of $c_{tt}$ versus $c_{ggh}$ (left) and $c_{tt}$ versus $c_{gghh}$ (right).
We observe that SM-like shapes (red) are preferred. However, doubly peaked structures are also possible for $\ctt$ values not too far from the SM value ($\ctt=0$). We also notice the similarity to Fig.~\ref{fig:HM2} (left).
The behaviour with respect to $\cgg$ is again similar.

Note that in SMEFT, $c_{ggh} $ and $c_{gghh}$ are related, so this behaviour would necessarily be the case. However we will see later that a shape classification algorithm based on unsupervised learning is able to detect shape differences which distinguish effects of $c_{ggh} $ and $c_{gghh}$.
An interesting feature is also that kind 1 (black) and kind 4 (blue) shapes appear only when we modify the value of $c_{hhh}$: for $c_{hhh}=1$
shapes of kind 1 never occur, and shapes of kind 4 are very unlikely.
Further, the kind 4 shapes tend to point to (moderately) negative values of $c_{hhh}$ as long as $\ctt$ is close to zero, as can be seen from Figs.~\ref{fig:HM1}, \ref{fig:HM2} and~\ref{fig:HM4}.


\subsection{Classification of $p_{T,h}$ distributions} 
\label{sec:pTanalysis}

\begin{figure}[htb!]
\centering
\includegraphics[width=0.75\linewidth]{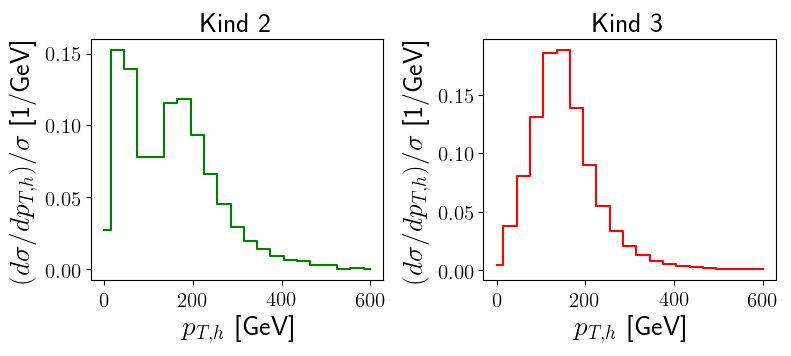}
  \caption{Examples of $\pth$ distributions with a single or double peak.}
  \label{fig:PT_KIND12}
\end{figure}

So far we have studied $\mhh$ distributions, assuming that they are very well suited to study the sensitivity to shape changes induced by anomalous couplings.
In order to verify that we do not miss out interesting features in the transverse momentum distributions, we also present a study of the $\pth$ distributions, 
but only at LO, to assess the salient features.
The main difference with respect to the $\mhh$ case is that in the $\pth$ analysis we could identify only two kinds of clearly distinct shapes: single peak (SM-like, which we denote as `$\pth$ kind 3') 
and double peak, with peaks separated by at least 30\,GeV (denoted as `$\pth$ kind 2'). Examples of such $\pth$ shapes are shown in Fig.~\ref{fig:PT_KIND12}.

\begin{figure}[htb!]
\centering
\includegraphics[width=0.75\linewidth]{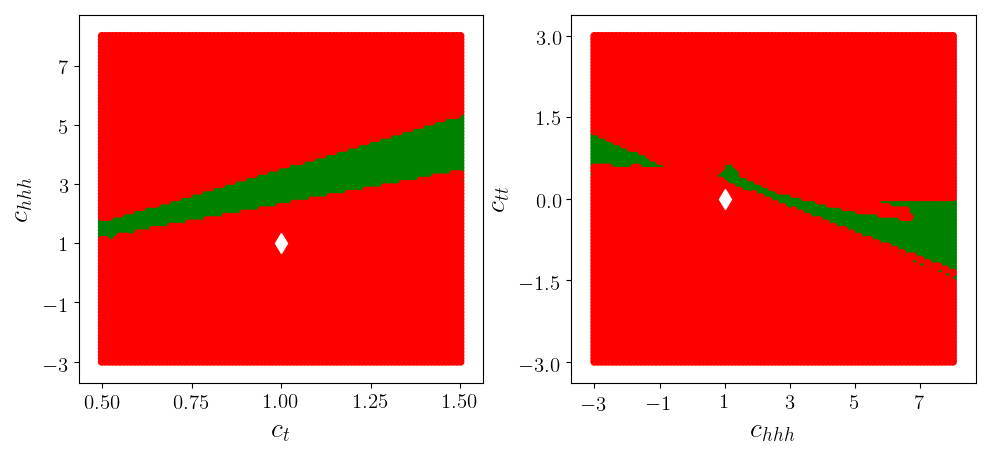}
  \caption{ The parameter space associated to each shape type in the $c_t-c_{hhh}$ and $\chhh-\ctt$ planes for the $\pth$ distribution.}
  \label{fig:KINDS_PT}
\end{figure}

The parameter spaces leading to singly or doubly peaked shapes are shown in Fig.~\ref{fig:KINDS_PT} for the $c_t-c_{hhh}$ and $\chhh-\ctt$ configurations.
The parameter region related to shapes with a well separated double peak (green) is similar to the $\mhh$ case, as one can see comparing with Figs.~\ref{fig:HM1} and~\ref{fig:HM2}. 
This indicates that the underlying parameter space leads to similar characteristics for the distributions differential in $\pth$ and $\mhh$, however the $\pth$ distribution is less sensitive than the $\mhh$  distribution.

\section{Classification and clustering by unsupervised learning}
\label{sec:unsupervised}


\subsection{Unsupervised learning procedure} 
\label{sec:unsupervised_procedure}

To assess the bias introduced by the definition of the four shape
types, and to find a more flexible classification  which
can be extended easily to more than four shape types, 
we  approach the classification problem using unsupervised learning techniques. 
We construct a classification of the shapes of the $\mhh$ distribution into distinct types, where we do not predefine what the types should look like.
For this purpose we use an autoencoder to find common patterns in the data and thus achieve a compressed representation.
The setup is implemented using {\sc Keras}~\cite{keras} and {\sc TensorFlow}~\cite{tensorflow}.
As input data we use 30 bins of width 20\,GeV for the normalised $\mhh$ distributions. 
We train the network based on a set of $10^5$ distributions, retaining 10\% for the validation. 
The encoder architecture, i.e. the part compressing the array information, is composed of two dense layers with 20 nodes and  
a middle layer with 4 nodes, the latter defining the length of the array containing the compressed information. 
The decoder architecture, which reconstructs the original array from
the compressed one, is composed of two dense layers of 20 nodes and an
output layer of the length of the input array.
We have also tried other encoder architectures, varying the number of
nodes in the layers as well as the number of layers, and found that the results deteriorate for less
than three layers. Adding more nodes had the tendency to lead to overfitting.

To test how stable our results are against variations of the training
data set and the encoding procedure, and to reduce uncertainties, for
example due to overfitting, 
we produced ten different autoencoder models. 
For each model we picked $10^4$ random points from the training set for validation
to start from  different training and validation sets and a different
initialization of the weights.
We trained the autoencoder for each model over 10000 epochs using {\tt
  Adam}~\cite{Kingma:2014vow} as optimizer and the root mean square
error to define the loss function.
Based on the trained autoencoder we applied the encoder models to the
training and validation data to obtain two sets of compressed arrays
for each of the ten models.
The ten different encoded training data sets are then fed to a
classification algorithm, where we employed the {\tt KMeans}
clustering algorithm from {\tt scikit-learn}~\cite{scikit}, asking for
a classification into a given number of clusters.
We tested classifications into four to eight clusters.

\begin{figure}[htb]
\centering
\includegraphics[width=0.9\linewidth]{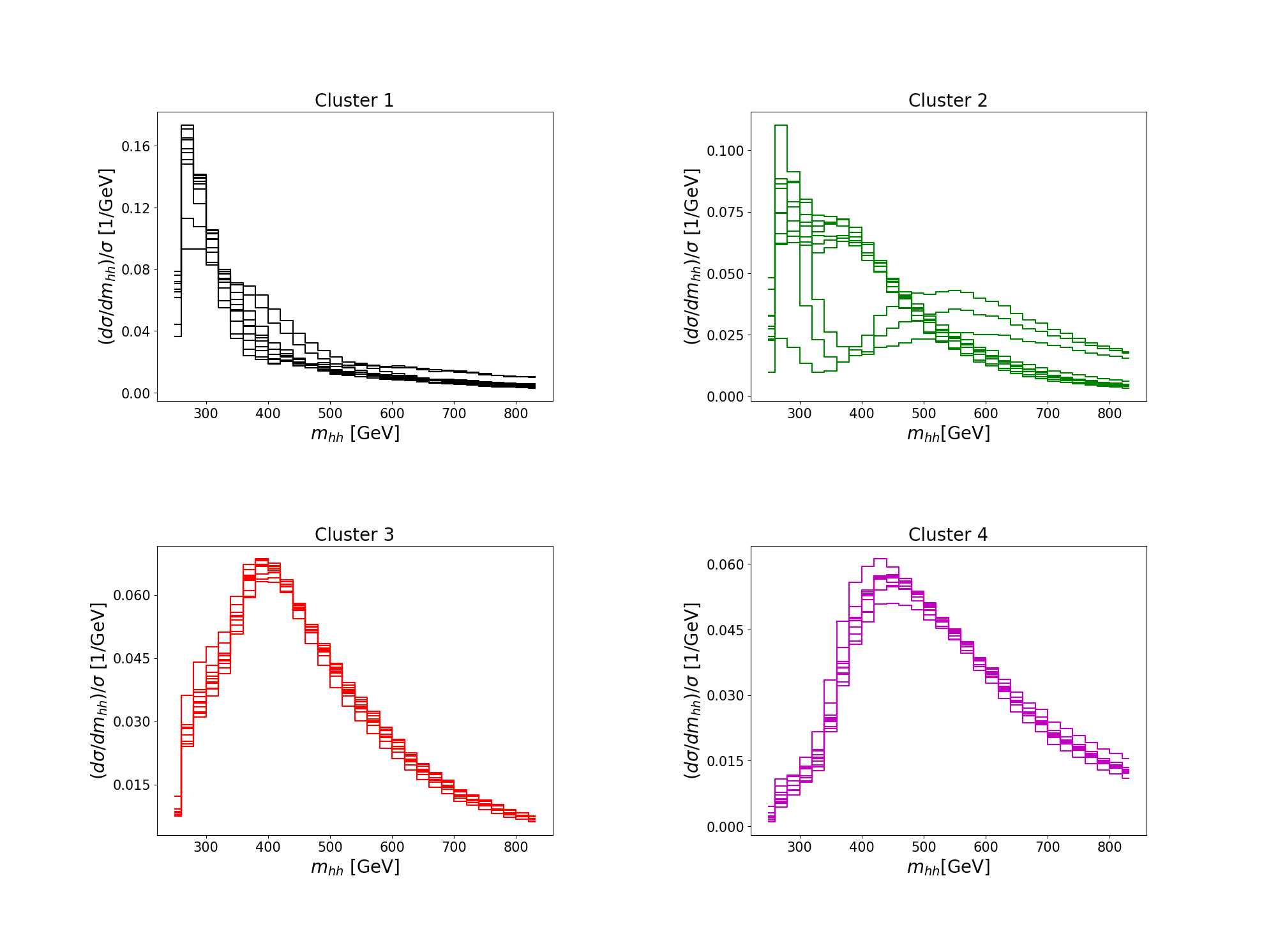}
  \caption{The clusters obtained by asking for a classification into
    four shape types. We show the cluster centres obtained from 10 different
    encoder models, in
    the colour code defined in Table~\ref{tab:clusters_colours}.}
  \label{fig:KMeans4}
\end{figure}

\begin{figure}[htb]
\centering
\includegraphics[width=1.2\linewidth]{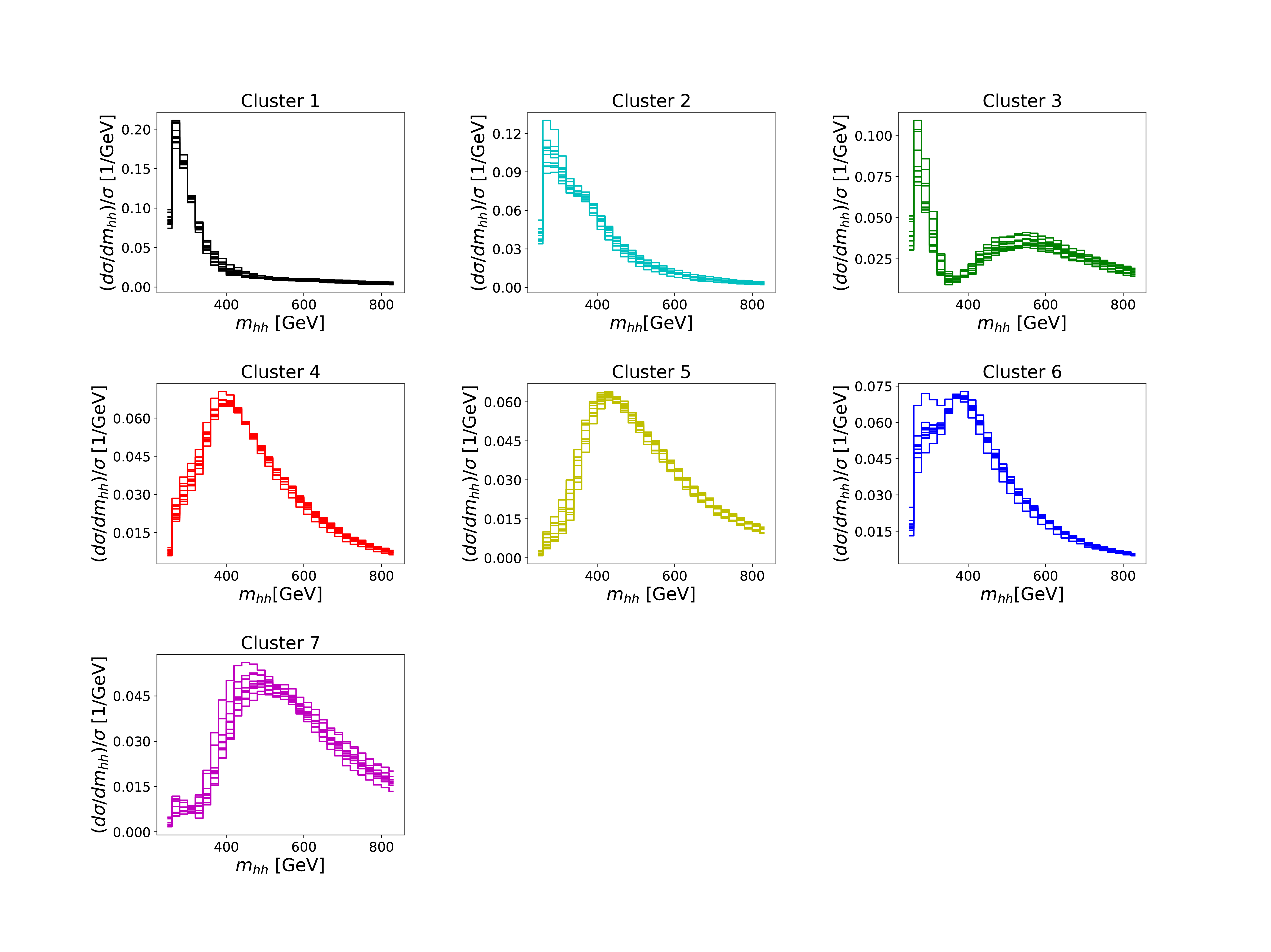}
  \caption{The clusters obtained by asking for a classification into
    seven shape types. The cluster centres obtained from 10 different
    encoder models are shown in
    the colour code defined in Table~\ref{tab:clusters_colours}.}
  \label{fig:KMeans7}
\end{figure}

Asking the {\tt KMeans} algorithm to find four clusters yielded the
shape types shown in Fig.~\ref{fig:KMeans4}, the result of asking for seven clusters
is shown in  Fig.~\ref{fig:KMeans7}.
The curves denote
the cluster centres determined by the  {\tt KMeans} algorithm, for
each of the ten encoder models, with a colour code as defined in Table~\ref{tab:clusters_colours}.

One can see from Figs.~\ref{fig:KMeans4} and \ref{fig:KMeans7} that in the case of clustering
into four shape types, cluster 2 contains shapes which vary substantially.
In contrast, for seven shape types, the cluster centers obtained
from the ten different encoder models are quite similar.
Asking for 5--8 clusters we found that seven clusters seemed to be the optimal number to
capture distinct shape features, while 
defining eight clusters did not lead to useful additional features but rather to the 
tendency to focus on local minima in the clustering space, while neglecting more gobal shape features.


The four clusters shown in Fig.~\ref{fig:KMeans4}  do only partly coincide with the ones defined in Section~\ref{sec:shapes}.
Shapes of kind 1, showing an enhanced $\mhh$ region, as well as shapes
of kind 3 (SM-like), were clearly identified. 
Shapes having a double peak were clustered together with shapes
showing a shoulder.
However, a cluster was formed which was not considered in the predefined types, containing shapes with an enhanced tail.

To combine the results from the ten clustering procedures,  we adopted
the ``majority vote'' method, i.e. for each of the ten
clustering procedures a given point in the coupling
parameter space gets a label (``vote'') corresponding to the cluster it belongs
to. The final cluster assigned to that point is the one which
collected the largest number of votes.

\subsection{Parameter space underlying the clusters} 

\begin{figure}[htb!]
  \centering
\includegraphics[width=0.8\linewidth]{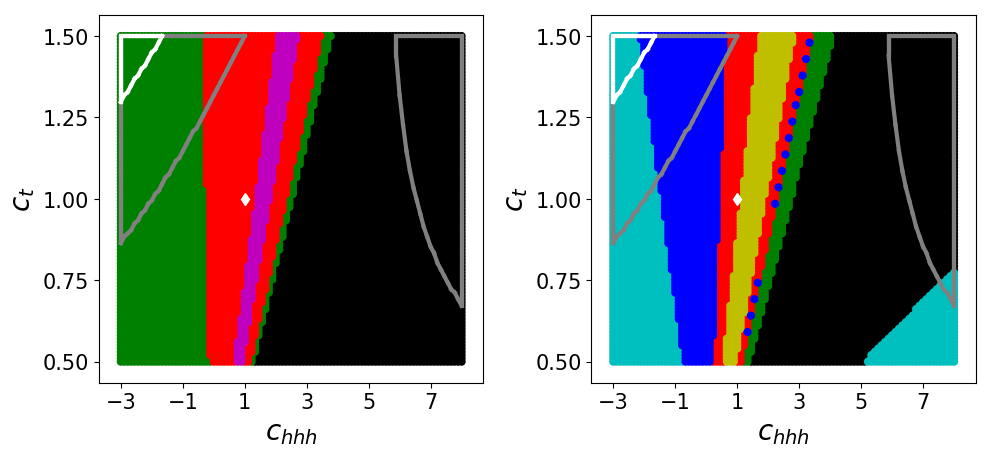}
\includegraphics[width=0.8\linewidth]{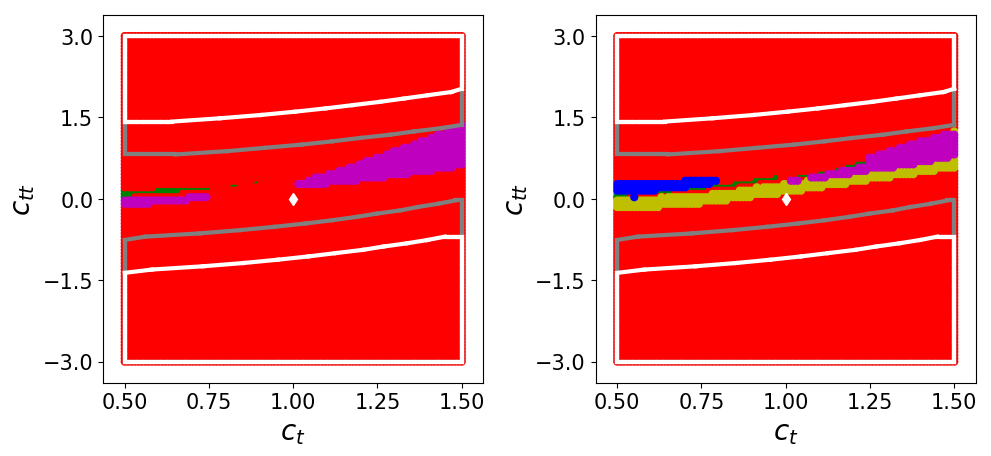}
  \caption{Shape types produced by variations of $\ct$ versus $\chhh$ (top) and  $\ct$ versus $\ctt$ (bottom). 
  Left: 4 clusters, right: 7 clusters.
  The areas outside the silver and white curves are regions where
  the total cross section exceeds $6.9\times \sigma_{SM}$ and $22.2 \times \sigma_{SM}$, respectively.
These values are motivated by the current ATLAS/CMS limits at $\sqrt{s}=13$\,TeV~\cite{Aad:2019uzh,Sirunyan:2018two}.
The red areas denote SM-like shapes.
The full colour code is given in Table \ref{tab:clusters_colours}.}
  \label{fig:ct_c3}
\end{figure}

\begin{figure}[htb!]
  \centering
\includegraphics[width=0.8\linewidth]{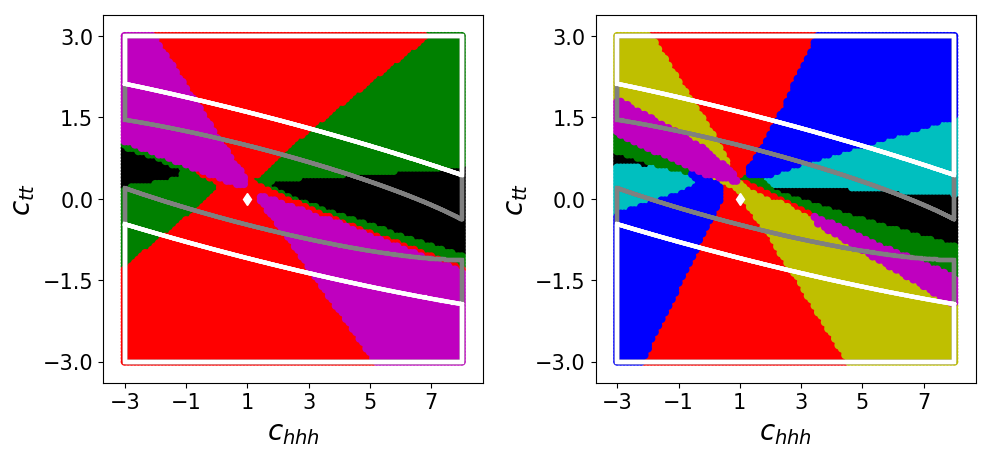}
\includegraphics[width=0.8\linewidth]{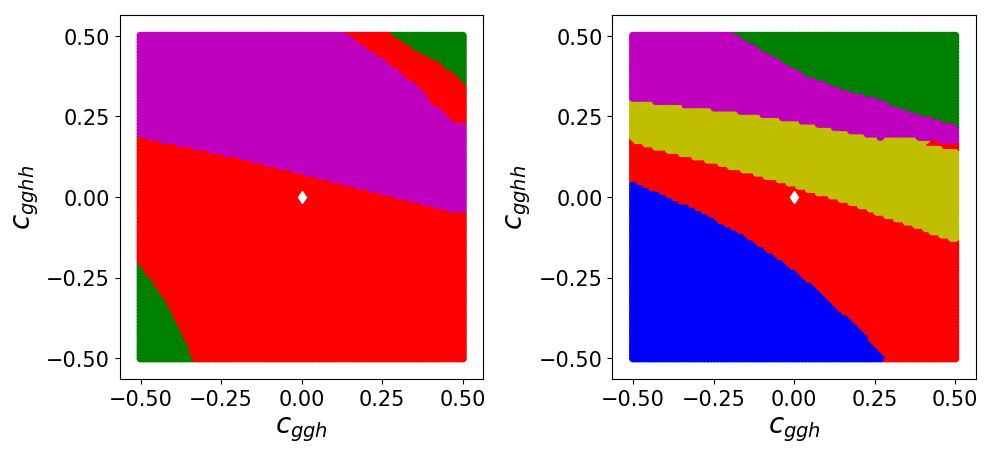}
  \caption{Shape types produced by variations of $\chhh$ versus $\ctt$ (top) and $\cg$ versus $\cgg$ (bottom). Left: 4 clusters, right: 7 clusters. The areas outside the silver and white curves are regions where
  the total cross section exceeds $6.9\times \sigma_{SM}$ and $22.2 \times \sigma_{SM}$, respectively.
These values are motivated by the current ATLAS/CMS limits at $\sqrt{s}=13$\,TeV~\cite{Aad:2019uzh,Sirunyan:2018two}. The red areas denote SM-like shapes. The full colour code is given in Table \ref{tab:clusters_colours}.}
  \label{fig:cg_cgg}
\end{figure}

\begin{figure}[htb!]
  \centering
\includegraphics[width=0.8\linewidth]{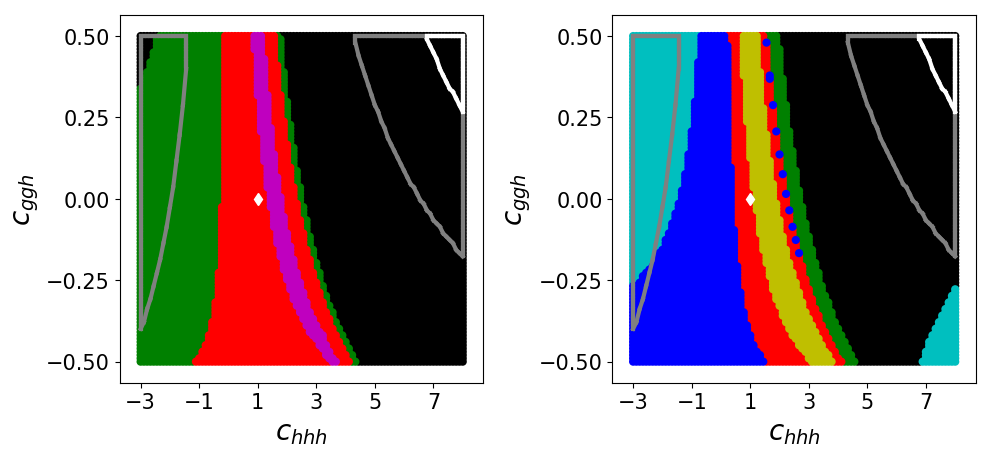}
\includegraphics[width=0.8\linewidth]{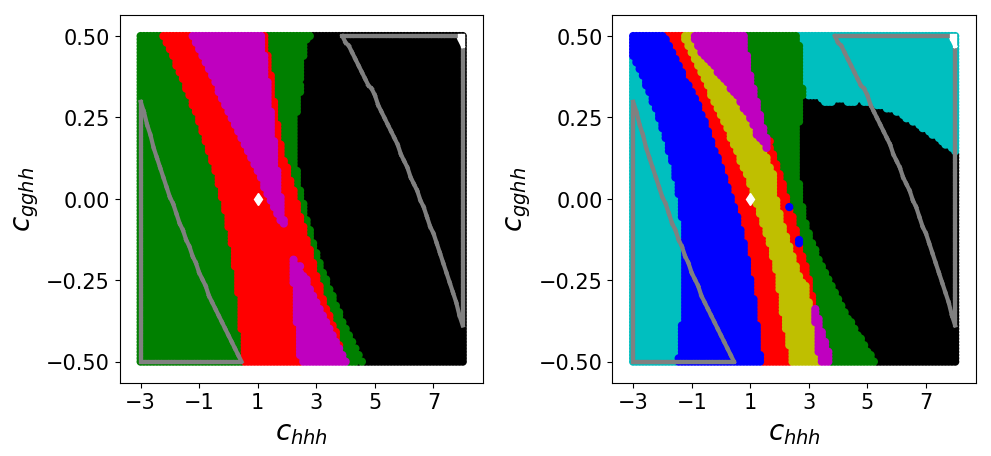}
  \caption{Shape types produced by variations of $\chhh$ versus $\cg$
    (top) and $\cgg$ (bottom). Left: 4 clusters, right: 7 clusters. 
  The areas outside the silver and white curves are regions where
  the total cross section exceeds $6.9\times \sigma_{SM}$ and $22.2 \times \sigma_{SM}$, respectively.
These values are motivated by the current ATLAS/CMS limits at $\sqrt{s}=13$\,TeV~\cite{Aad:2019uzh,Sirunyan:2018two}.
The red areas denote SM-like shapes.
  The full colour code is given in Table \ref{tab:clusters_colours}.}
  \label{fig:c3_cgghh}
\end{figure}

\begin{figure}[htb!]
  \centering
\includegraphics[width=0.8\linewidth]{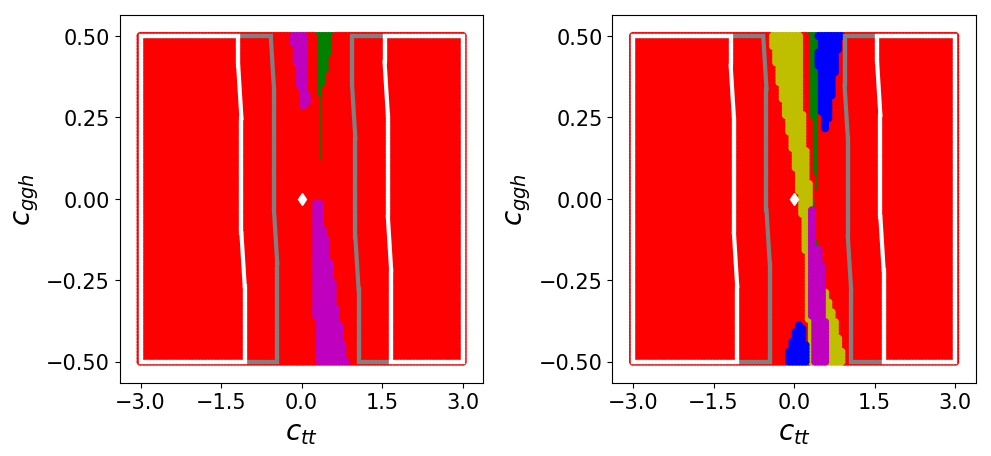}
\includegraphics[width=0.8\linewidth]{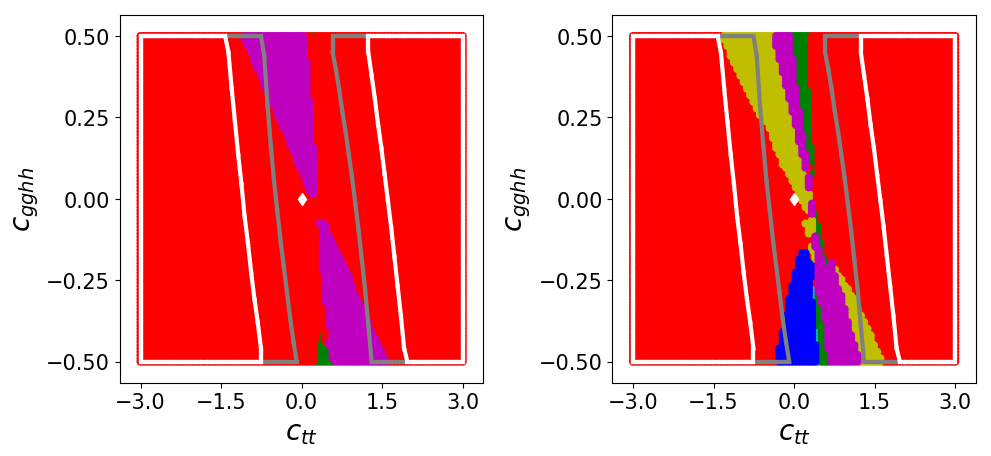}
  \caption{Shape types produced by variations of $\ctt$ versus $\cg$
    (top) and $\cgg$ (bottom). Left: 4 clusters, right: 7 clusters. The areas outside the silver and white curves are regions where
  the total cross section exceeds $6.9\times \sigma_{SM}$ and $22.2 \times \sigma_{SM}$, respectively.
These values are motivated by the current ATLAS/CMS limits at $\sqrt{s}=13$\,TeV~\cite{Aad:2019uzh,Sirunyan:2018two}.
The red areas denote SM-like shapes.}
  \label{fig:ctt_cgg}
\end{figure}

\begin{figure}[htb!]
  \centering
\includegraphics[width=0.8\linewidth]{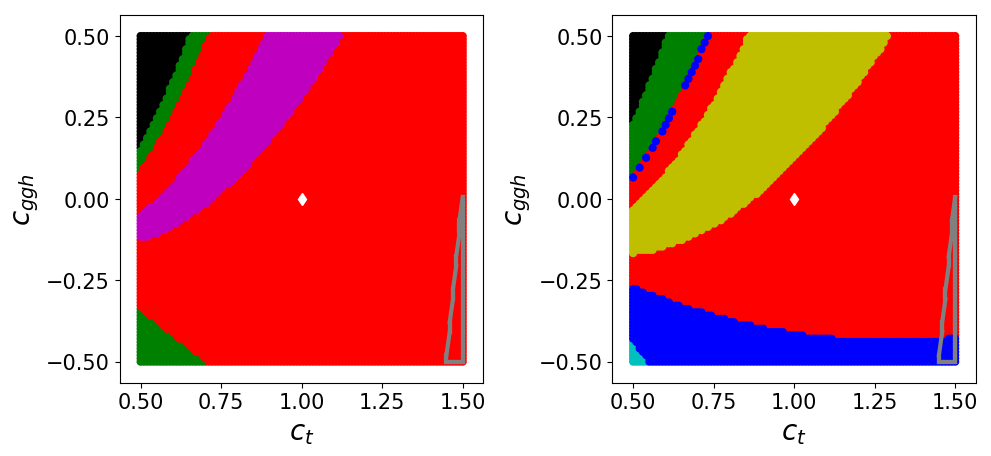}
\includegraphics[width=0.8\linewidth]{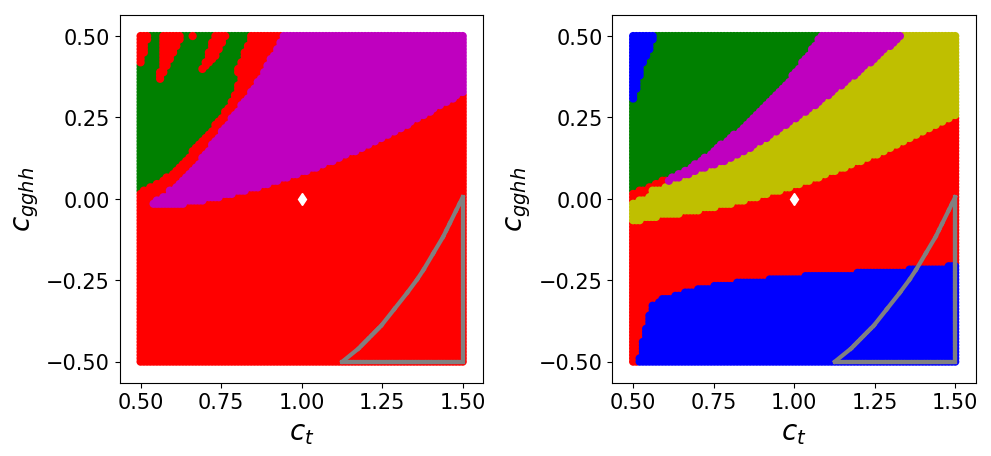}
  \caption{Shape types produced by variations of $\ct$ versus $\cg$
    (top) and  $\cgg$ (bottom). Left: 4 clusters, right: 7 clusters. The areas outside the silver and white curves are regions where
  the total cross section exceeds $6.9\times \sigma_{SM}$ and $22.2 \times \sigma_{SM}$, respectively.
These values are motivated by the current ATLAS/CMS limits at $\sqrt{s}=13$\,TeV~\cite{Aad:2019uzh,Sirunyan:2018two}.
The red areas denote SM-like shapes.}
  \label{fig:ct_cgg}
\end{figure}

In this section we show how the parameter space relates to the clusters if we ask for four or seven clusters. 
For each parameter configuration of our 5-dimensional grid, we plot the corresponding cluster type in Fig.~\ref{fig:ct_c3} and Figs.~\ref{fig:cg_cgg} to~\ref{fig:ct_cgg}. 
\begin{table}[htb]
\begin{center}
\begin{tabular}{|c| c | c |}
\hline
 Cluster&closest predefined type&colour\\
\hline
\hline
\multicolumn{3}{|c|}{4 clusters}\\
\hline
\hline
 1& kind 1 (enhanced low $\mhh$)&  black \\
 2& double peak/enhanced tail &  magenta \\
 3& kind 3 (SM-like)&  red \\
 4& kind 4 (close-by double peaks)/shoulder&  blue \\
\hline
\hline
\multicolumn{3}{|c|}{7 clusters}\\
\hline
\hline
 1& enhanced low $\mhh$ &  black \\
 2& enhanced low $\mhh$, slowly falling or shoulder&  cyan\\
 3& enhanced low $\mhh$, second local maximum above  $\mhh\simeq 2m_t$&  green \\
 4& SM-like &  red \\
 5& SM-like with enhanced tail & yellow \\
 6& close-by double peaks or shoulder left&  blue \\
 7& no steep slope at low $\mhh$, enhanced tail&  magenta \\
\hline
\end{tabular}
\end{center}
\caption{Clusters and shape types with corresponding colour codes for the classification into four and seven clusters.}
\label{tab:clusters_colours}
\end{table}
The colour codes are shown in Figs.~\ref{fig:KMeans4} and~\ref{fig:KMeans7}, and are also listed in Table~\ref{tab:clusters_colours}.
For clusters which are similar to the shape types defined in Section~\ref{sec:shapes}, 
we should also find patterns similar to the ones shown in Figs.~\ref{fig:HM1} to \ref{fig:HM5}.

Comparing  Fig.~\ref{fig:ct_c3} (top row) with Fig.~\ref{fig:HM1} (left), both showing variations of $\ct$ versus $\chhh$, we see that kind 1 shapes (black)  are clearly identified.  
However, for both four and seven clusters the area for SM-like shapes
got smaller, as the clustering algorithm  also identifies features which were not considered in the predefined shapes.
For example, the clustering into four clusters identifies shapes which are almost SM-like but have an enhanced tail (magenta), 
and the clustering into seven clusters in addition identifies shapes
which are almost SM-like but have a shoulder (blue).
Certainly we could have defined such features in our analyzer as well,
but it is not that easy to define where the tail starts and what
exactly should be considered as ``enhanced''.
Further, the figure clearly shows that small variations of $\chhh$ can easily distort the SM-like shape, while the shape is more robust against variations of $\ct$.
Fig.~\ref{fig:ct_c3} (bottom row) shows $\ct$ versus $\ctt$. We again see that variations $\ct$ and $\ctt$  mostly produce SM-like shapes. 
Why this is so can be understood from the behaviour of the coefficients $A_i$ in eq.~(\ref{eq:Ai_mhh}) which are relevant in these cases.
For Fig.~\ref{fig:ct_c3} (top row), only the coefficients $A_1,A_3$ and $A_7$ are relevant. 
As $A_1$ and $A_7$ have opposite signs and a different peak location, this can generate a rich shape structure.
For Fig.~\ref{fig:ct_c3} (bottom row), the coefficients  $A_2,A_6$ and $A_8$ are relevant in addition to $A_1,A_3$ and $A_7$.
$A_2$ being the coefficient of $\ctt^2$, it is dominant except for
very small values of $\ctt$ and leads to a SM-like shape.
We also observe that $\ctt$ has the tendency to enhance the total
cross section, such that only a relatively small slice in $\ctt$ is
left after considering the bounds on the total cross section.
\begin{figure}[htb!]
    \centering
\includegraphics[width=0.49\linewidth]{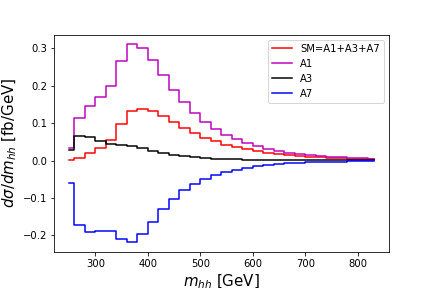}
\includegraphics[width=0.49\linewidth]{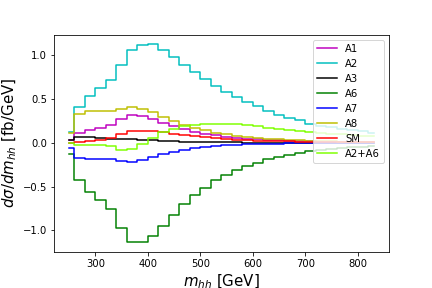}
  \caption{Contributions of the coefficients $A_i$ in eq.~(\ref{eq:Ai_mhh}) which are relevant for Fig.~\ref{fig:ct_c3}.}
  \label{fig:Ai_ct_c3_ctt_}
\end{figure}
%

Fig.~\ref{fig:cg_cgg} (top row) shows $\chhh$ versus $\ctt$,
where we see that the interplay between $\chhh$ and $\ctt$ can lead to all shape types.
Comparing Fig.~\ref{fig:cg_cgg} (bottom row) with Fig.~\ref{fig:HM1} (right), showing variations of $\cg$ versus $\cgg$, we observe that 
the unsupervised learning algorithm with seven clusters distinguishes four shape types, showing that
large values of $\cg$ and $\cgg$ favour shapes with an enhanced tail (magenta) or/and a double peak (green), while negative values 
favour a shoulder on the left of the peak (blue).
The limits on the total cross section do not exclude any parameter range in this panel.

A behaviour similar to the one in Fig.~\ref{fig:ct_c3} can be seen in Fig.~\ref{fig:c3_cgghh}:
as $\chhh$ varies the disribution goes through various shape types, while variations of $\cg$ and $\cgg$ affect the shapes to less extent.
Fig.~\ref{fig:c3_cgghh} also shows that a positive $\cgg$ value has the tendency to enhance the tail of the distribution.

Fig.~\ref{fig:ctt_cgg} shows $\ctt$ versus $\cg$ (top) and $\ctt$ versus $\cgg$ (bottom). 
Compared to Fig.~\ref{fig:HM5}, 
the clustering into both four and seven clusters shows a better
discrimination power between SM-like shapes and small deviations, for
example due to an enhanced tail.
We again see that $\ctt$ has a larger impact on the total cross
section than $\cg$ or $\cgg$.

Fig.~\ref{fig:ct_cgg}, showing the $\ct-\cg$ and $\ct-\cgg$ parameter planes, can be compared to Fig.~\ref{fig:HM3}. Again, both the case of four and of seven clusters  
indicates that the unsupervised learning algorithm is able to distinguish better subtle influences on the shape than our method based on humanly classified shapes.

\begin{figure}[htb]
  \centering
\includegraphics[width=0.8\linewidth]{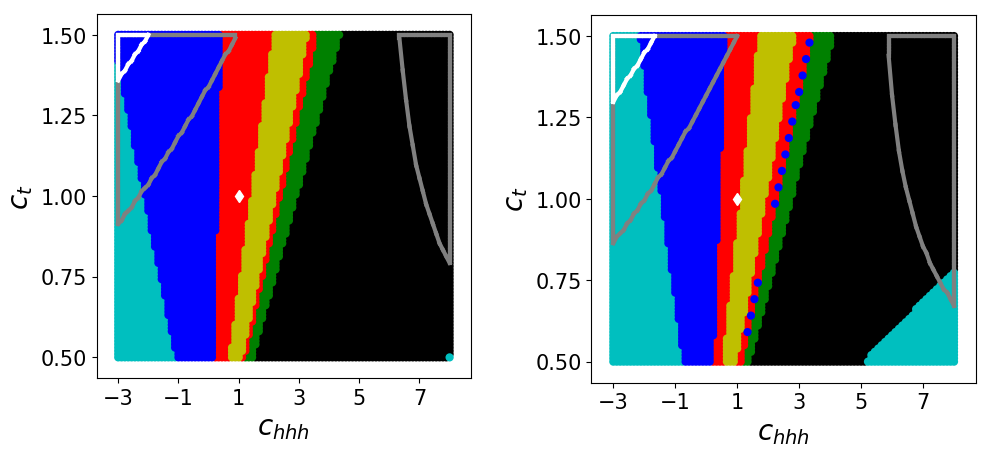}
  \caption{Comparison of LO and NLO results for shape types produced
    by variations of $\ct$ versus $\chhh$ Left: LO, right: NLO.}
  \label{fig:LO_NLO}
\end{figure}

In Fig.~\ref{fig:LO_NLO}, we compare results of our shape
analysis produced with LO and NLO input data.  We observe that NLO
corrections can change the shape considerably and therefore are
important for a shape analysis.

\begin{figure}[htb]
  \centering
\includegraphics[width=0.6\linewidth]{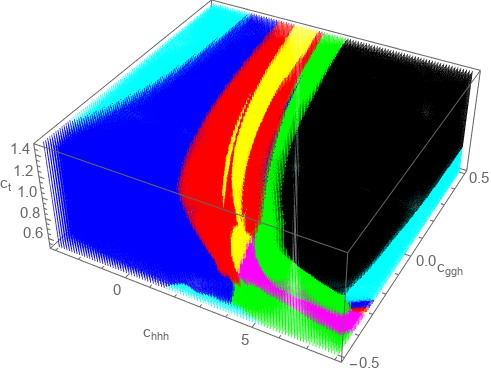}
  \caption{Three-dimensional visualisation of shape types produced
    by variations of $\ct$, $\chhh$ and $\cg$ simulating the SMEFT situation. For $\cgg$ the value
    given by Eq.~(\ref{cghh}) has been used, for $\ctt$ we used $\ctt=0.05\ct$.}
  \label{fig:SMEFT}
\end{figure}

The results above have shown that the parameters $\chhh$ and $\ctt$
have the largest influence on the shape.
In SMEFT,  $\ctt$ is suppressed compared to $\ct$ by one order of the large new physics
scale~\cite{DiMicco:2019ngk}.
Furthermore, SMEFT imposes the relation (\ref{cghh}) between $\cg$ and
$\cgg$.
Using this relation and imposing that $\ctt$ amounts to 5\% of $\ct$,
we obtain a 3-dimensional parameter space simulating the SMEFT
situation, which is visualized in
Fig.~\ref{fig:SMEFT}.

\clearpage

\subsection{Identification of benchmark points}

\begin{figure}[htb]
  \centering
\includegraphics[width=1.1\textwidth]{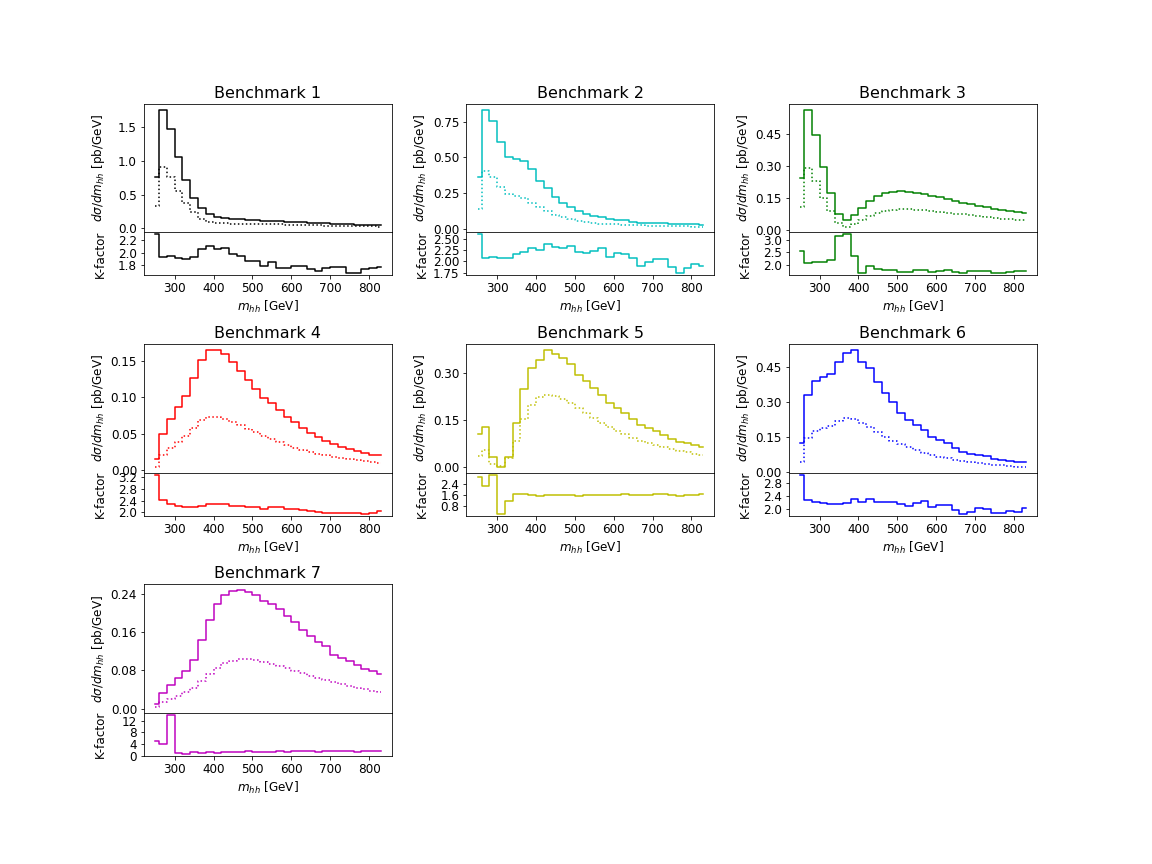}
  \caption{Higgs boson pair invariant mass distributions
corresponding to the benchmark points listed in
Table~\ref{tab:benchmarks}. The solid curves denote the NLO result,
the dotted curves the LO result. The lower panels show the K-factor,
defined as $d\sigma_{\rm{NLO}}/d\sigma_{\rm{LO}}$.}
  \label{fig:benchmarks}
\end{figure}

The determination of the cluster centres allows to identify NLO benchmark
points which should be representative for each characteristic shape.
As the cluster centres determined by the  {\tt KMeans} algorithm do
not necessarily correspond to grid points of our input grids in
parameter space, and as we work with normalised distributions to find
the cluster centres, 
we determine the benchmark points based on the
following procedure
\begin{enumerate}
\item We identify the grid point in parameter space corresponding to a
  curve which is closest to the cluster centre, where the distance
  measure is the bin-wise geometric distance to the cluster center of the encoded
  distribution.
\item If the corresponding total cross section exceeds the limit of
  $6.9\times \sigma_{SM}$~\cite{Aad:2019uzh}, we proceed to the curve
  which is the next-closest to the cluster center.
  \item If several curves determined this way have an identical distance
    measure, we choose the one where the value of $\ct$ is closer to
    the SM value (anticipating that the top-Higgs Yukawa coupling will
    be constrained increasingly well from other processes).
  \item If after this procedure there are still several curves
    satisfying these criteria, from these curves we pick the one closest to the cluster
    center according to  the Kullback–Leibler~\cite{Kullback-Leibler}
  distance measure, applied to the normalised distributions.

\end{enumerate} 
Following this procedure we find the benchmark points listed in Table~\ref{tab:benchmarks}.
In Fig.~\ref{fig:benchmarks} we show the $\mhh$ distributions
corresponding to these benchmark points, at LO as well as at NLO.

\begin{table}
\begin{tabular}{|c|c|c|c|c|c|r|c|c|}
\hline
benchmark & $c_{t}$ & $c_{hhh}$ & $c_{tt}$ & $c_{ggh}$ & $c_{gghh}$ & $\sigma_{\rm{NLO}}$ [pb] & K-factor & ratio to SM \\
\hline
1 & 0.94 &  3.94 &  -$\frac{1}{3}$ & 0.5 &  $\frac{1}{3}$ & 182.50 $\pm$ 5.11 & 1.93 & 6.64 \\
\hline
2 & 0.61 & 6.84 & $\frac{1}{3}$ &  0.0 & -$\frac{1}{3}$ & 135.63 $\pm$ 4.27 & 2.16 & 4.93 \\
\hline
3 & 1.05 &  2.21 &  -$\frac{1}{3}$ & 0.5 & 0.5 & 109.24  $\pm$ 2.65 & 1.86 & 3.97 \\
\hline
4 & 0.61 &  2.79 &  $\frac{1}{3}$ &  -0.5 & $\frac{1}{6}$ &50.44 $\pm$ 1.53 & 2.16 & 1.83 \\
\hline
5 & 1.17 &  3.95 &  -$\frac{1}{3}$ & $\frac{1}{6}$ &  -0.5 & 116.68 $\pm$ 6.25 & 1.63 & 4.24 \\
\hline
6 & 0.83 &  5.68 &  $\frac{1}{3}$ &  -0.5 &  $\frac{1}{3}$ & 145.37 $\pm$ 8.25 & 2.19 & 5.29 \\
\hline
7 & 0.94 & -0.10 &  1& $\frac{1}{6}$ &  -$\frac{1}{6}$ &  96.69 $\pm$ 1.45 & 2.29 & 3.52 \\
\hline
\end{tabular}
\label{tab:benchmarks}
\caption{NLO benchmark points derived from the cluster centers as
  described in the text.}
\end{table}

\clearpage

\clearpage

\section{Conclusions}

The aim of this work was to provide more insight how certain configurations of anomalous couplings in the Higgs sector lead to a corresponding characteristic shape of the Higgs boson pair invariant mass distribution. 
For this purpose we employed the Lagrangian relevant to Higgs boson pair production as given in a non-linear Effective Field Theory framework, which contains five (potentially) anomalous couplings~\cite{Buchalla:2018yce}.
We produced data for the Higgs boson pair invariant mass distribution, based on a calculation which includes the NLO QCD corrections with full top quark mass dependence,
varying all five coupling parameters by finite steps, thus producing a dense grid of data.
Then we defined four characteristic shape types for the $\mhh$
distribution and visualised the parameter space leading to these shape types.
To this aim we projected onto all possible two-dimensional slices of the parameter
space, keeping the remaining parameters at their Standard Model values.
We also considered $\pth$ distributions for a shape analysis, however we found that the $\mhh$ distribution is more sensitive to shape changes induced by anomalous couplings.

Further, we tested an unsupervised learning approach to classify shapes.
We produced $10^5$ distributions, trained a neural network based on an autoencoder to extract common shape features and tried to find the number of shape clusters which optimally catches different  shape characteristics. 
Our study demonstrated that some shape features, like an enhanced tail or a shoulder in the $\mhh$ distribution, were caught very well by this procedure, 
and provided more insight about the underlying parameter space leading
to such features than
the analysis based on predefined shape classes.
While machine learning is not essential to define shape clusters, it
has the advantage of being easily extendible to a different number of
shape types, different binnings or other observables,  and of minimising the human bias compared to other shape analysis methods.

The shape analysis revealed that the Standard-Model-like shape is quite stable against variations of $\ct,\cg$ and $\cgg$, as long as $\chhh=1$, while deviations of $\chhh$ from the SM value show a rich shape changing pattern.
We also found that small deviations of $\ctt$ from zero are very
likely to produce a doubly peaked structure in the $\mhh$
distribution, while SM-like shapes dominate again as $\ctt$ moves
further away from zero. However, as $\ctt$ leads to a rather fast
increase of the total cross section, the shape analysis in combination with
the limits on the total cross section allows to put constraints on $\ctt$.
This is an interesting feature because, in contrast to $\ct$ and $\cg$, $\ctt$ cannot be constrained directly from single Higgs boson processes.
Further, an enhanced tail or a shoulder of the $\mhh$ distribution are
likely to be produced by nonzero values of $\cgg$, however the
influence of the effective Higgs-gluon couplings on the shape is milder than the one of $\chhh$ and $\ctt$.

We also provide seven benchmark points, based on
the full NLO calculation and taking into account current experimental constraints, which lead to the characteristic shapes as
represented by our cluster centers.

The method can also be applied to other processes where anomalous couplings introduce characteristic shape changes to differential cross sections, and it  can be extended to consider more than one distribution simultaneously.

\section*{Acknowledgements}
We would like to thank Gerhard Buchalla, Alejandro Celis, Long Chen, Victor Diaz, Stephan Jahn, Stephen P.~Jones, Matthias Kerner, Gionata Luisoni, Ludovic Scyboz and Johannes Schlenk for collaboration on the HH code, for useful discussions and for valuable comments on the manuscript.
This research was supported in part by the COST Action CA16201 (`Particleface') of the European Union.

\bibliographystyle{JHEP}

\bibliography{ML_main}

\end{document}